\documentclass[aps,prb,twocolumn,showpacs,amsmath,amssymb]{revtex4}
\usepackage[colorlinks=true,citecolor=blue,linkcolor=red,linktocpage=true
]{hyperref}
\usepackage{graphics}
\usepackage{epsfig}

\begin{document}
   \title{
Spectroscopy of electron flows with single- and two-particle emitters
         }
\author{ Michael Moskalets$^{1,2,3}$ and Markus B\"uttiker$^{1}$}
\affiliation{$^1$D\'epartement de Physique Th\'eorique, Universit\'e de Gen\`eve, CH-1211 Gen\`eve 4, Switzerland \\$^2$Institut f\"{u}r Theorie der Statistischen Physik, RWTH Aachen University, D-52056 Aachen, Germany \\$^3$Department of Metal and Semic. Physics, NTU "Kharkiv Polytechnic Institute", 61002 Kharkiv, Ukraine}
\date\today
   \begin{abstract}
To analyze the state of injected carrier streams of different electron sources, we propose to use correlation measurements at a quantum point contact with the different sources connected via chiral edge states to the two inputs. 
In particular we consider the case of an on-demand single-electron emitter correlated with the carriers incident from a biased normal reservoir, a contact subject to an alternating voltage and a stochastic single electron emitter. 
The correlation can be viewed as a spectroscopic tool to compare the states of injected particles of different sources. 
If at the quantum point contact the amplitude profiles of electrons overlap, the noise correlation is suppressed. In the absence of an overlap the noise is roughly the sum of the noise powers due to the electron streams in each input. We show that the electron state emitted from a (dc or ac) biased metallic contact is different from a Lorentzian amplitude electron state emitted by the single electron emitter (a quantum capacitor driven with slow harmonic potential), since with these inputs the noise correlation is not suppressed.
In contrast, if quantized voltage pulses are applied to a metallic contact instead of a dc (ac) bias then the noise can be suppressed. We find a noise suppression for multi-electron pulses and for the case of stochastic electron emitters  for which the appearance of an electron at the quantum point contact is probabilistic.
\end{abstract}
\maketitle

\section{Introduction}
\label{intro}

The experimental realization \cite{Feve07} of an on-demand, high-frequency, single-electron source (SES) makes it  possible to inject into a solid state circuit single particles, electrons and holes, in a controllable way.
Using several uncorrelated single-electron sources mesoscopic circuits were proposed which permit to vary the amount of fermionic-correlations \cite{OSMB08} and to produce controllably orbitally entangled pair of particles \cite{SMB09}. 
Similar high-frequency sources of single electrons were realized using dynamical quantum dots without \cite{Blumenthal07} or with a perpendicular magnetic field \cite{Wright08}. 
The principal advantage of on-demand, single-electron sources over the usually used metallic contacts (MCs) as electron sources is the possibility, in the former case, to switch on and off quantum correlations between particles initially emitted from uncorrelated sources.
An example of correlations generated by normal metallic contacts is the two-particle Aharonov-Bohm effect in the solid state Hanbury Brown-Twiss interferometer discussed theoretically \cite{YS92,SSB04,SNB09} and found experimentally \cite{Neder07}. 
In contrast, with single-electron sources the two-particle interferometer, as it is discussed in Ref.\,\onlinecite{SMB09}, can show or not show the Aharonov-Bohm effect depending on whether sources are driven in synchronism or not.

\begin{figure}[b]
  \vspace{0mm}
  \centerline{
   \epsfxsize 8cm
   \epsffile{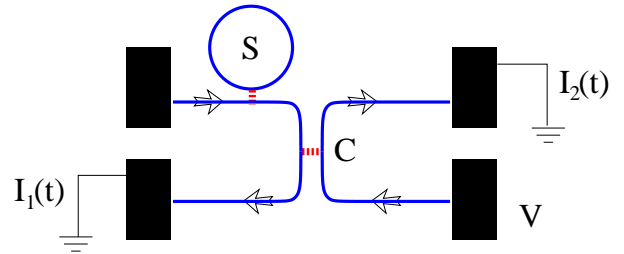}
             }
  \vspace{0mm}
  \nopagebreak
\caption{(color online)
A mesoscopic electron collider circuit with a single-electron source $S$, a circular edge state, and a metallic contact source biased with a voltage $V$. At the quantum point contact $C$ the particles emitted by the two sources can collide if the times of emission are adjusted properly. Solid blue lines are edge states with direction of movement indicated by arrows. Short dashed red lines are quantum point contacts connecting different parts of a circuit. Black rectangles are metallic contacts.
}
\label{fig1}
\end{figure}

The appearance of quantum correlations (fermionic, in the case of electrons) between initially uncorrelated particles is due to the overlap of wave-packets on the wave splitter.  
For electrons in solid state circuits the splitter is a quantum point contact (QPC), (see Fig.\,\ref{fig1}, the QPC labeled $C$). 
Such correlations are well known in optics, see, e.g., Ref.\,\onlinecite{Mandel99}. The overlap of fermions was discussed in Ref.\,\onlinecite{Loudon91} and in Ref.\,\onlinecite{BB00}. 
The overlap depends on the spatial extend of wave-packets and also on the times when they arrive at the wave splitter. 
Thus the resulting correlations can be used to access information about the space-time extend of quantum states. For MCs working as electron sources such information is rather hidden since the mentioned correlations are always present. In contrast with on-demand single-electron sources control of the emission time can be achieved,  i.e. the appearance or disappearance of correlations can be controlled.  Thus with such sources the space time extend of quantum states becomes accessible.

In mesoscopics physics shot noise \cite{BB00} is the natural quantity that can be used to find information on two-particle correlations.  \cite{Buttiker90,Samuelsson06,Beenakker06} The shot noise of carriers emitted by two SES's is suppressed if the wave-packets overlap at the QPC connecting edge states in which emitted particles propagate. \cite{OSMB08} If two sources are identical and they emit particles at the same time, the emitted particles are in identical quantum states and the shot noise is suppressed down to zero. This effect is similar to the Hong-Ou-Mandel effect in optics \cite{HOM87} with the evident difference that electrons are rather forced go into different output channels while photons are bunched into the same output channel.

The aim of this paper is to use the shot noise suppression as a spectroscopic tool allowing comparison of the quantum states emitted by the different electron sources.
As the test state we will use the one emitted by the SES. 
The SES is made of a quantum capacitor \cite{Feve07} in the quantum Hall effect regime. 
The SES is connected to one of the arms of the mesoscopic electron collider, Fig.\,\ref{fig1}.
Under the action of a potential $U(t) = U\cos (\Omega t)$ periodic in time  the SES emits a sequence of alternating electrons and holes. \cite{note1} 
In a certain range of amplitudes, in the quantized emission regime, the SES emits one electron and one hole. 
At low driving frequency, in the adiabatic regime, the emitted state by the SES is close to the state generated by voltage pulses of Lorentzian form with a time integral equal to a flux quantum: such a quantized voltage pulse produces a single-particle state on top of the Fermi sea. \cite{ILL97,KKL06,ZSAM09}.
In the second arm we put the source of interest and investigate the resulting shot noise.

The paper is organized as follows:
In Sec.\,\ref{sec1} we calculate the zero-frequency cross-correlator of currents flowing into two outputs in the collider circuit with a SES in one input and a biased metallic contact in other one.
In Sec.\,\ref{sec2} we address the effect of stochastic single-particle emitters which emit or do not emit a particle in a given period that arrives at the QPC.  We demonstrate  the irrelevance of such stochastic emission to the shot noise suppression effect.
In Sec.\,\ref{sec3} the shot noise suppression effect is found for colliding single- and two-electron pulses.
A discussion of our results is given in Sec.\,\ref{concl}. Much of the analysis is grouped into three Appendices. 
In Appendix\,\ref{app1} we present a detailed model of a single electron source.
In Appendix\,\ref{app2} we calculate a current correlation function for a periodically driven mesoscopic scatterer connected to reservoirs biased with periodic voltages.
In Appendix\,\ref{app3} the zero-frequency cross-correlation function for a circuit with two SESs is expressed it terms of single- and two-particle probabilities.

\section{Single electron emitter and biased metallic contact as an electron source}
\label{sec1}

We consider an electron collider with a SES in one branch and a MC with a potential  $V_{2}^{(\sim)}(t) = V_{2}^{(\sim)}(t + {\cal T})$ periodic in time in another branch, see, Fig.\,\ref{fig1}.
The potential $U(t)$ driving the SES and $V_{2}^{(\sim)}(t)$ have the same period, ${\cal T} = 2\pi/\Omega$, which is assumed to be large enough to consider adiabatic transport 
and neglect relaxation and decoherence processes \cite{FDJ08,DGF09} relevant for high-energy excitations.
In addition the MC is biased with a constant potential $V_{2}$ with respect to the other contacts which all have the same chemical potential $\mu$. 
The temperature is taken to be zero.

We utilize the scattering matrix approach \cite{BM10} to transport in mesoscopic systems and describe this circuit with the help of the frozen scattering matrix 

\begin{equation}
\hat S(t) = \left(
\begin{array}{cc}
e^{ikL_{1S}} S^{SES}(t) r_C & e^{ikL_{1V}} t_C \\
\ & \\
e^{ikL_{2S}} S^{SES}(t) t_C & e^{ikL_{2V}} r_C
\end{array}
\right) ,
\end{equation}
\ \\
\noindent
with $S^{SES}(t)$ the scattering amplitude of the SES, see Appendix \ref{app1}, Eq.\,(\ref{07_nsc_03_1}), $r_C/t_C$ the reflection/transmission amplitude for the central quantum point contact $C$, and $L_{j X}$ the length from the SES ($X = S$) or the MC ($X = V$) to the contact $j = 1,2$ where the corresponding current $I_{j}(t)$ is measured.
At zero temperature we need all quantities at the Fermi energy $\mu$ only.

We are interested in the zero-frequency correlation function \cite{BB00} ${\cal P}_{12}$ of the currents $I_{1}(t)$ and $I_{2}(t)$ flowing into the contacts $1$ and $2$, see Fig.\,\ref{fig1}.
The corresponding calculations are presented in Appendix \ref{app2}.
In the adiabatic regime and at zero temperature we have, ${\cal P}_{12} \equiv {\cal P}_{12}^{(sh,ad)}$, Eq.\,(\ref{ac15_1}):

\begin{equation}
{\cal P}_{12} = - {\cal P}_{0} \sum\limits_{ q = - \infty }^{\infty} \left| \left\{ S^{SES} \Upsilon_{2}^{*}  \right\}_{q} \right|^2\, \left| \dfrac{eV_{2} }{\hbar\Omega} - q \right|   \,,
\label{01}
\end{equation}
\ \\
\noindent
where ${\cal P}_{0} = e^2 R_{C} T_{C}/ {\cal T}$ is the shot noise\cite{OSMB08} produced by one particle (either an electron or a hole) emitted by the SES during the period ${\cal T}$.
The oscillating potential at contact $2$ appears in the form of a phase factor 
\begin{equation}
\label{01mb} 
\Upsilon_{2}(t)  = e^{-\,i\,  \frac{e}{\hbar} \int\limits_{-\infty}^{t} dt'\, V_{2}^{(\sim)}(t')  } \,,
\end{equation}
which multiplies the scattering amplitude of the SES. The symbol $\{..\}_q$ indicates the q-th Fourier component (in time) of these two amplitudes.  Eq. (\ref{01}) illustrates that the correlation tests the coherence properties of the two sources. 

Note the decoherence processes, which we neglect in the present work, can lead to suppression of coherence

\subsection{Quantized voltage pulse}

With a voltage pulse of Lorentzian shape and a time integral quantized to a single flux quantum~\cite{ILL97, KKL06} one can excite an electron from a Fermi sea without any other disturbance to the Fermi sea.
The state for an excited electron has a Lorentzian density profile (the time-dependent current is a Lorentzian pulse) which is similar to the one \cite{OSMB08} emitted by the SES in the adiabatic regime.
Thus we  can expect a shot noise suppression effect if an electron excited out of a metallic contact with a quantized voltage pulse and an electron emitted by the SES collide at the central QPC.
Below we show that this is really the case.

Thus let us assume that a periodic pulsed potential is applied to the MC,

\begin{eqnarray}
eV_{2}(t)   & = &  \dfrac{ 2\hbar\Gamma }{ \left(t - t_{0} \right)^2 + \Gamma^2  }\,,  \quad 0 < t \leq {\cal T} \,,
\label{ac17}
\end{eqnarray}
\ \\
\noindent
where $\Gamma \ll {\cal T}$ is the half-width of the  pulse, and $t_{0}$ is the time when the electron is excited.
Such a pulse excites one electron during the period ${\cal T}$.

The potential $V_{2}(t)$, Eq.\,(\ref{ac17}), has a dc component, $eV_{2} = h / {\cal T}$, and a component which is periodic in time,
\begin{eqnarray}
eV_{2}^{(\sim)}(t)   & = & \dfrac{ 2\hbar\Gamma }{ \left(t - t_{0} \right)^2 + \Gamma^2  } - \dfrac{h }{\cal T } \,, \quad 0 < t \leq {\cal T}\,.
\label{ac18}
\end{eqnarray}
\ \\
\noindent
The corresponding phase factor $\Upsilon_{2}( t)$, Eq.\,(\ref{01mb}),  (for $0 < t \leq {\cal T}$),

\begin{eqnarray}
\Upsilon_{2}(t) &=& e^{i \left\{\Omega t   - 2 \left( \arctan \frac{t - t_{0} }{\Gamma } + \arctan \frac{ t_{0} }{\Gamma } \right) \right\} } \,.
\label{ac19}
\end{eqnarray}
\ \\
\noindent

\begin{figure}[b]
  \vspace{0mm}
  \centerline{
   \epsfxsize 7.8cm
   \epsffile{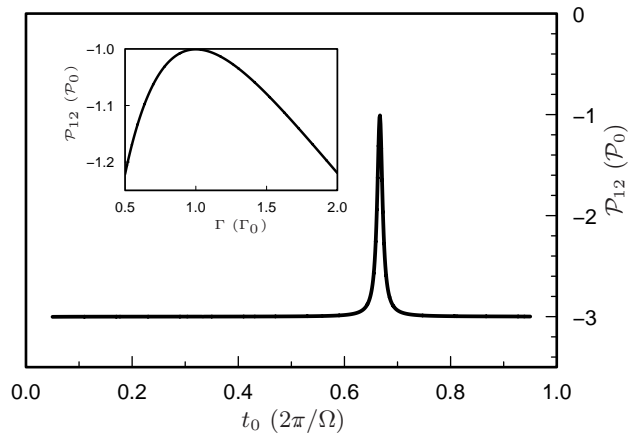}
            }
\begin{picture}(0,0)(0,0)  \put(-20,0){$t_{0}~(2\pi/\Omega)$} \put(117,80){\rotatebox{90}{${\cal P}_{12}~({\cal P}_{0}) $}}  \put(-35,75){\tiny{ ${\ \Gamma~(\Gamma_{0}) }$} } \put(-87,105){\rotatebox{90}{ \tiny{${\cal P}_{12}~({\cal P}_{0}) $} } } \end{picture}.
  \vspace{0mm}
  \nopagebreak
\caption{Main: The noise ${\cal P}_{12}$, Eq.\,(\ref{01}), as a function of the time $t_{0}$ when an electron is excited out of the metallic contact by the voltage pulse $V_{2}(t)$. The plot assumes that the half-width of the voltage pulse at the MC and the pulse of the SES are the same ($\Gamma = \Gamma_{0}$ ).  The parameters of the SES, Eq.\,(\ref{08_01}), are: $T = 0.1$, $U_{0} = 0.25$, $U_{1} = 0.5$. Inset: The noise at the minimum as a function of $\Gamma$.
}
\label{fig4_2}
\end{figure}
The result of a numerical evaluation of the shot noise based on Eq.\,(\ref{01}), as a function of the time $t_0$ is given in Fig.\,\ref{fig4_2}. For almost all times $t_0$ the noise is $-3{\cal P}_{0}$ and only at a very special co-incidence time is there a sharp reduction of the noise. 
The noise, $-3{\cal P}_{0}$, is produced by three uncorrelated particles emitted during a period: One electron is emitted by the metallic contact and two particles, an electron and a hole, are emitted by the SES.
However, if the MC and the SES emit electrons at the same time, $t_{0} = t_{0}^{(-)}$, then after colliding at the central QPC these electrons become correlated and do not contribute effectively to the shot noise.
The remaining value, $-{\cal P}_{0}$, is due to  the hole emitted by the SES.

The shot noise suppression (ShNS) effect depends on the overlap of wave-packets in time (hence the times of emission should be the same) and in space (hence the width of wave-packets should be the same).
If the wave packets have different width, see inset to Fig.\,\ref{fig4_2}, there is some extra noise.
Therefore, the ShNS effect provides a direct tool to compare the states of particles emitted from the sources of different types, not only from the similar sources.  The ShNS of two SES's was already discussed in Ref.~\onlinecite{OSMB08}.

The most used source of electrons in mesoscopics is a biased (with dc or ac voltage) metallic contact. 
Now we show that the electron collider circuit with SES and a biased MC as an electron source does not show what we call the shot noise suppression effect.
Therefore, the state of electrons emanating from a biased MC is different from the state of an electron emitted by the SES.

\subsection{DC bias}

If no ac bias is applied, $V_{2}^{(\sim)}(t) = 0$, the phase factor is $\left| \Upsilon_2(t) \right|^2 = 1$ and only the Fourier coefficients $\left| S^{SES}_{q}\right|^2 = 4\Omega^2\Gamma_{0}^2 \exp(-2\Omega\Gamma_{0} |q|)$ enter Eq.\,(\ref{01}). We recall that we assume an adiabatic limit $\Omega\Gamma_{0} \ll 1$, where $2\Gamma_{0}$ is the time during which an electron (a hole) is emitted by the SES.
Since $\left| S_{q}\right|^2 = \left| S_{-q}\right|^2$, we conclude from Eq.\,(\ref{01}) that in this case the shot noise is independent of the sign of the voltage.
Therefore, the result will be the same no matter whether the MC emits electrons, $eV > 0$, or holes, $eV < 0$.
For definiteness we will use $eV_{2} > 0$.

Evaluation of the cross-correlator gives 
\begin{equation}
{\cal P}_{12}   = - {\cal P}_{0} \left\{ \dfrac{eV_{2} }{\hbar\Omega}  + 2 e^{- 2\Omega\Gamma_{0} \left[ \frac{eV_{2} }{\hbar\Omega} \right] } \left(1 + 2\Omega\Gamma_{0} \left[ \frac{eV_{2} }{\hbar\Omega} \right] \right)  \right\} .
\label{02}
\end{equation}
\ \\
\noindent
where we have introduced the integer part $[eV_{2} /(\hbar\Omega)].$
This correlator has the following asymptotics,

\begin{equation}
{\cal P}_{12} \,=\, \left\{
\begin{array}{ll}
-\, 2\, {\cal P}_{0}\,, & eV_{2} \ll \hbar\Omega\,, \\
\ \\
-\, (e^3 V_{2}/h ) R_CT_C \,, & eV_{2} \gg \hbar \Gamma_{0}^{-1} \,.
\end{array}
\right.
\label{08}
\end{equation}
\ \\
\noindent
Here the first line is the shot-noise due to the SES emitting one electron and one hole during the period.
The second line is the shot noise due to a dc biased metallic contact alone. \cite{BB00}
The latter noise is due to scattering at the quantum point contact $C$ of extra electrons flowing out of a biased contact above the Fermi sea with chemical potential $\mu$.
These electrons are emitted with rate $eV_{2}/h$.
Therefore, one could naively expect that if the rate of emission of electrons from the SES and from the MC is the same, $\hbar\Omega = eV_{2}$, then each emitted electron will collide at the central QPC with an electron propagating within another edge state and the shot noise gets suppressed.

This is not the case! As follows from Eq.\,(\ref{02}), the shot noise has no strong feature at $eV_{2} \sim \hbar\Omega$. The shot noise is a monotonous function of the dc bias $V_{2}$.
A possible reason for this is that the states of electrons emitted from the SES and from the dc biased MC are quite different:
The electrons emitted from the SES can be thought as wave-packets with spatial extend proportional to the duration of emission $\Gamma_{0}$.
In contrast the electrons emitted by the metallic contact are rather plane-wave like extended along the whole edge state.
Thus their overlap at the central QPC is minute hence they do not acquire any significant correlations.
The shot noise remains roughly the sum of the noises produced independently be the SES and by the dc biased MC.

Next we show that the noise suppression effect is also absent if the metallic contact is driven by an ac bias.

\subsection{AC bias}

\begin{figure}[b]
  \vspace{0mm}
  \centerline{
   \epsfxsize 8cm
   \epsffile{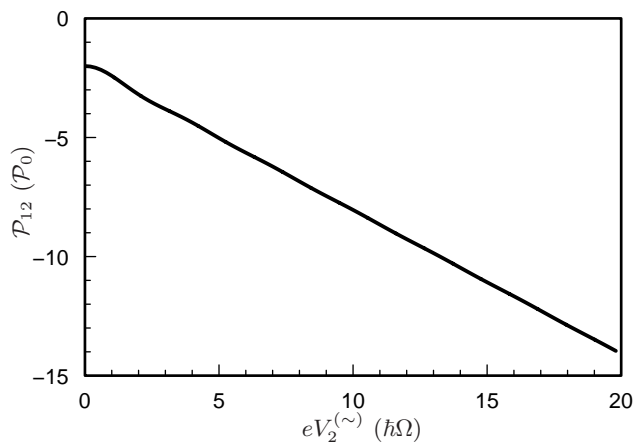}
            }
\begin{picture}(0,0)(0,0)  \put(-10,0){$eV_{2}^{(\sim)}~(\hbar \Omega)$} \put(-120,75){\rotatebox{90}{${\cal P}_{12}~({\cal P}_{0}) $}} \end{picture}
  \vspace{0mm}
\nopagebreak
\caption{
The noise ${\cal P}_{12}$, Eq.\,(\ref{01}), as a function of the amplitude $eV_{2}^{(\sim)}$ of the ac potential applied to the metallic contact. The parameters of the single electron source in Eq.\,
(\ref{08_01}), are: $T = 0.1$, $U_{0} = 0.25$, $U_{1} = 0.5$.
}
\label{fig4_1}
\end{figure}

Consider next the case of a metallic contact with an ac bias, 
$V_{2}^{(\sim)}(t) = V_{2}^{(\sim)} \cos(\Omega t)$, $V_{2} = 0$.
In this case the metallic contact emits both electrons and holes.
The cross-correlator, Eq.\,(\ref{01}), as a function of amplitude $V_{2}^{(\sim)}$ is given in Fig.\,\ref{fig4_1}.
There is a small feature at $eV_{2}^{(\sim)} = \hbar\Omega$ visible in Fig.\,\ref{fig4_1} but this feature is minute compared to the huge dip of interest in this work.
Therefore, there is no indication of a shot noise suppression at $eV_{2}^{(\sim)} \sim \hbar\Omega$ when the rate of emission of particles from the SES and from the MC is the same.

This is in contrast to the case when sinusoidal voltages are applied to both inputs \cite{vrmpmb}. Then the discussion can best be cast into excitations of electron-hole pairs \cite{mmmb} which create a shot noise which has been measured \cite{reydellet} . For two oscillating voltages theory predicts significant two-particle correlations due to the Hanbury Brown-Twiss effect which depend on the phase delay of the two oscillating voltages \cite{vrmpmb}.

One can wonder whether 
the absence of the shot noise suppression effect is possibly due to fluctuations in emission of electrons from the biased metallic contact.
Our expectation is that neither fluctuations nor a possible presence of multi-electron (multi-hole) states play a crucial role.
To show it we consider next two circuits.

\section{Shot-noise suppression effect with stochastic single-electron sources}
\label{sec2}

\begin{figure}[b]
  \vspace{0mm}
  \centerline{
   \epsfxsize 8cm
   \epsffile{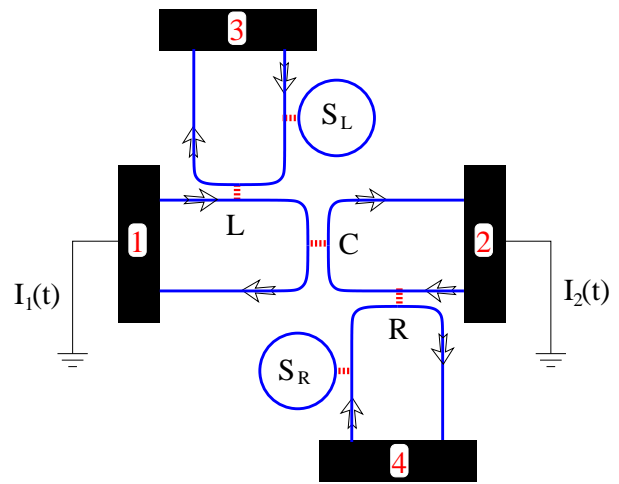}
             }
  \vspace{0mm}
  \nopagebreak
\caption{(color online)
A mesoscopic electron collider circuit with two stochastic single-particle streams originated from the quantum point contacts $L$ and $R$.
In the case two particles enter the central part of a circuit they can collide
at the quantum point contact $C$ if the times of emission by the sources $S_{L}$ and $S_{R}$ were adjusted properly.
Solid blue lines are edge states with direction of movement indicated by arrows.
Short dashed red lines are quantum point contacts connecting different parts of the circuit.
Black rectangles numbered by $1$ to $4$ are metallic contacts. }
\label{fig2}
\end{figure}

In Fig.\,\ref{fig2} we show a circuit with two single-electron sources, $S_{L}$ and $S_{R}$ each emitting one electron and one hole per period ${\cal T}$.
Initially the particle stream is regular.
However, say for particles emitted by the source $S_{L}$, at the quantum point contact $L$ an electron (a hole) can be either reflected to the metallic collector $3$ or be transmitted to the central part of a circuit.
Thus, the single-electron source $S_{j}$ together with a corresponding quantum point contact $j = L, R$ comprise a stochastic single-electron source which can either inject into the central part of a collider one electron (hole) during a given period or not.

We assume that all the metallic contacts are grounded and calculate the zero temperature cross-correlator ${\cal P}_{12} \equiv {\cal P}_{12}^{(sh,ad)}$, Eq.\,(\ref{ac15_1}),

\begin{equation}
{\cal P}_{12} = \dfrac{e^2\Omega}{4\pi}\, \sum\limits_{q = -\infty}^{\infty} |q| \sum\limits_{\gamma, \delta = 1}^{4} \left\{S_{1\gamma}\, S_{1\delta}^{*} \right\}_{q}\, \left\{S_{2\gamma}\, S_{2\delta}^{*} \right\}_{q}^{*}\,.
\label{n01}
\end{equation}
\ \\
\noindent
Since there is no bias all the phase factors are $\Upsilon_{\delta} = 1$ in Eq.\,(\ref{ac14c}) and $V_{\gamma\delta} = 0$ in Eq.\,(\ref{ac15_1}).
The elements of the frozen scattering matrix are expressed in terms of the transmission/reflection amplitudes $t_{i}/r_{i}$ for the quantum point contacts $i = L, R, C$,  time-dependent amplitudes $S_{j}^{SES}(t)$ for sources $S_{j}$, $j = L, R$, and corresponding phase factors $e^{ikL_{\alpha\beta}}$ with $L_{\alpha\beta}$ a length between metallic contacts $\alpha$ and $\beta$.
For instance, $S_{13}(t) = e^{i k L_{13}} S_{L}^{SES}(t) t_{L} r_{C}$.
For $S_{j}^{SES}(t)$ we use Eq.\,(\ref{07_nsc_03_1}) with emission times $t_{0}^{(\pm)}$ and a pulse half-width $\Gamma_{0}$ replaced by $t_{j}^{(\pm)}$ and $\Gamma_{j}$. respectively.
Then we find:

\begin{eqnarray}
{\cal P}_{12}   & = & - 2{\cal P}_{0} \bigg\{\left(T_{L} - T_{R} \right)^2 \nonumber \\
\label{n02} \\
&& + T_{L}T_{R} \left[\gamma\left(\Delta t^{(-)} \right) + \gamma\left(\Delta t^{(+)} \right) \right] \bigg\}\,, \nonumber
\end{eqnarray}
\ \\
\noindent
where $\Delta t^{(\pm)} = t_{L}^{(\pm)} - t_{R}^{(\pm)}$ with $t_{j}^{(\pm)}$ ($j = L, R$) the time of an electron ($-$)/hole ($+$) emission by the SES $j$, and the suppression function

\begin{eqnarray}
\gamma\left(\Delta t \right) &=& \dfrac{\left(\Delta t \right)^2 + \left(\Gamma_{L} - \Gamma_{R} \right)^2 }{ \left(\Delta t \right)^2 + \left(\Gamma_{L} + \Gamma_{R} \right)^2 }\,,
\label{n03}
\end{eqnarray}
\ \\
\indent
If the single electron sources emit particles at different times, $\Delta t^{(\pm)} \gg \Gamma_{L}, \Gamma_{R}$, then the correlation is, 

\begin{equation}
{\cal P}_{12}   =  - 2{\cal P}_{0} \left\{ T_{L}^{2} + T_{R}^2 \right\} .
\label{n02_01}
\end{equation}
\ \\
\noindent
This expression is due to the shot noise produced by the four uncorrelated particles (two electrons and two holes) emitted by the two sources during the period ${\cal T}$.
Apparently the single-particle contribution (to the cross-correlator) is negative.
We call this regime {\it classical}, since the shot noise can be explained in terms of single-particle probabilities only, see Appendix~\ref{app3}..

On the other hand, if the pulses of the same width, $\Gamma_{L} = \Gamma_{R}$, are emitted at the same time, $t_{L}^{(-)} = t_{R}^{(-)}$ (for electrons) and  $t_{L}^{(+)} = t_{R}^{(+)}$ (for holes), then the cross-correlator is suppressed:

\begin{equation}
{\cal P}_{12}   =  - 2{\cal P}_{0} \left(T_{L} - T_{R} \right)^2    .
\label{n02_02}
\end{equation}
\ \\
\noindent
If in addition the circuit is symmetric, $T_{L} = T_{R}$, then the cross-correlator is suppressed down to zero.

This suppression is due to a positive two-particle contribution arising (in addition to negative single-particle contributions which are also present)  when particles (either two electrons or two holes) collide at the quantum point contact $C$.
Due to such collisions each of the particles loses information about its origin (i.e., about the source that emitted it) and the pair of particles propagating to contacts $1$ and $2$ in Fig.\,\ref{fig2} becomes orbitally entangled. \cite{SMB09}
We call this regime {\it a quantum regime}, since to describe a shot noise we additionally need to take into account the existence of both direct and exchange two-particle quantum mechanical amplitudes for colliding particles, see Appendix \ref{app3}, Eq.\,(\ref{07_ppa_09}).

Thus with this circuit we showed that the shot noise suppression effect is sensitive to a space-time confinement of electron states rather than to a regularity in appearance of electrons at the place (the QPC $C$) where they can overlap.

\section{Shot-noise suppression effect with single- and two-particle sources}
\label{sec3}

\begin{figure}[b]
  \vspace{0mm}
  \centerline{
   \epsfxsize 8cm
   \epsffile{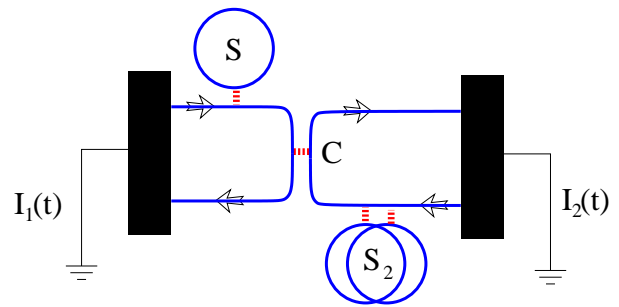}
             }
  \vspace{0mm}
  \nopagebreak
\caption{(color online)
A mesoscopic electron collider circuit with a single-electron source $S$ and a two-particle source $S_{2}$.
At the quantum point contact $C$ the particles emitted by different sources can collide if the times of emission were adjusted properly.
Solid blue lines are edge states with direction of movement indicated by arrows.
Short dashed red lines are quantum point contacts connecting different parts of a circuit.
Black rectangles are metallic contacts.
}
\label{fig3}
\end{figure}

Now we consider a circuit (see Fig.\,\ref{fig3}) which contains both a single particle emitter $S$ and a two-particle emitter $S_2$. As a two-particle source we use two single-electron sources placed close to each other and emitting in synchronism. \cite{SOMB08}
In the adiabatic case of interest here the scattering amplitude $S^{TES}(t)$ is the product of scattering amplitudes of single-electron sources comprising a two-particle source.
For simplicity we assume both sources to be identical.
Then $S^{TES}$ is the square of the amplitude given by Eq.\,(\ref{07_nsc_03_1}) with $t_{0}^{(\pm)}$ and $\Gamma_{0}$ replaced with $t_{2}^{(\pm)}$ and $\Gamma_{2}$, respectively.
At the time $t_{2}^{(-)}$ ($t_{2}^{(+)}$) the pair of electrons (holes) is emitted by the source $S_{2}$.
The cross-correlator ${\cal P}_{12}$, Eq.\,(\ref{ac15_1}), reads

\begin{equation}
{\cal P}_{12} = - {\cal P}_{0} \sum\limits_{q = -\infty}^{ \infty} |q| \left|\left\{  S^{SES} \left( S^{TES}\right)^{*} \right\}_{q} \right|^{2} ,
\label{n04}
\end{equation}
\ \\
\noindent
where $S^{SES}(t)$ is the scattering amplitude, Eq.\,(\ref{07_nsc_03_1}), for a single-electron source $S$ and $S^{TES}(t)$ is the scattering amplitude for the two-electron (two-particle) source $S_{2}$ shown in Fig.\,\ref{fig3}.

Simple calculations yield:

\begin{eqnarray}
{\cal P}_{12}   & = & - {\cal P}_{0} \bigg\{ \gamma^{2}\left(\Delta t^{(-)}\right) + 2 \gamma\left(\Delta t^{(-)} \right) + \chi\left(\Delta t^{(-)} \right) \nonumber \\
\label{n05} \\
&& +   \gamma^{2}\left(\Delta t^{(+)}\right) + 2 \gamma\left(\Delta t^{(+)} \right) + \chi\left(\Delta t^{(+)} \right) \bigg\}\,.
\nonumber
\end{eqnarray}
\ \\
\noindent
where $\Delta t^{(\pm)} = t_{0}^{(\pm)} -  t_{2}^{(\pm)}$.
The function $\chi(\Delta t)$ is,

\begin{equation}
\chi(\Delta t)   =  \dfrac{ 16\Gamma_{2}^{2}\Gamma_{0}^{2} }{\left((\Delta t)^2 + \left(\Gamma_{2} + \Gamma_{0} \right)^{2} \right)^{2} } \,,
\label{m_02_14}
\end{equation}
\ \\
\noindent
and the suppression function $\gamma(\Delta t)$ is given in Eq.\,(\ref{n03}) with $\Gamma_{L}$ and $\Gamma_{R}$ replaced by $\Gamma_{0}$ (for a SES) and $\Gamma_{2}$ (for a two-particle source), respectively.

If all the particles are emitted at different times, $\Delta t^{(\pm)} \gg \Gamma_{0}, \Gamma_{2}$, the cross-correlator, ${\cal P}_{12} = - 6 {\cal P}_{0}$, is due to contributions of six uncorrelated particles (three electrons and three holes) emitted during the period ${\cal T}$.
While for simultaneous emission, $\Delta t^{(\mp)} = 0$, the cross-correlator is partially suppressed.
If $\Gamma_{2} = \Gamma_{0}$, the cross-correlator is suppressed down to the level due to two-particles, ${\cal P}_{12} = - 2{\cal P}_{0}$.
So when the two-electron wave-packet collides with a single-electron wave-packet, two colliding electrons, one from each side, produce no noise while the remaining electron produces noise as if it propagated alone through the QPC.
The same holds for hole wave-packets.

\section{Conclusion}
\label{concl}

A method to compare quantum states of initially uncorrelated electrons in mesoscopic circuits was proposed.
The electron streams should be directed onto a quantum point contact from different sides and the cross-correlator of currents flowing out of the QPC should be measured.
In general two uncorrelated streams produce additive noises.
However, if the particles overlap at the QPC they become correlated and the noise gets suppressed.
The closer the quantum states of particles resemble each other the better the overlap that can be achieved, hence the noise is suppressed more strongly.

We considered several sources of electrons, in particular, (i) a metallic contact, emitting a rather continuous stream of electrons with a rate proportional to the bias, and (ii) a periodically driven quantum capacitor, a single-electron source, emitting traveling wave-packets of electrons which are rather localized in space and alternate with wave packet of holes.
We found that the streams produced by the MC biased with a dc (ac) voltage and by the SES remain almost uncorrelated after passing the QPC even if the electrons are emitted with the same rate.
Therefore, we conclude that the electrons of these streams are in quite different quantum states.
On the other hand, if the periodic sequence of quantized voltage pulses is applied to an MC, then the resulting electron stream can be easily correlated with a stream emitted by the SES resulting in a complete suppression of the shot noise. From this we can conclude 
that the electrons of these streams are in the same quantum states

If the streams are fluctuating then the shot noise can be suppressed by the amount proportional to the average number of particles overlapping at the QPC. We also found a partial suppression of the shot noise in the case of pulses carrying different number of particles.
Basically the remaining noise is due to the difference of the numbers of particles carried by the colliding pulses.

\begin{acknowledgments}
We thank M. Albert, P. Degiovanni and G. F\`{e}ve for discussion and communication.
M. M. thanks the council of the doctoral school program of Western Switzerland for an invitation to present a series of lectures. 
M. B. is supported by the Swiss NSF, MaNEP, and the European Networks NanoCTM and Nanopower. 
\end{acknowledgments}

\appendix

\section{Scattering amplitude and current of a SES}
\label{app1}

As a single-electron source we use a quantum capacitor \cite{PTB96,Gabelli06}$^,$\cite{Feve07}$^,$\cite{MSB08} described by a model in which a single circular edge state of circumference $L$ in a cavity is  coupled via a quantum point contact (QPC) with transmission probability $T$ to a linear edge state (see the left upper corner of the Fig.\,\ref{fig1}).
A potential $U(t) = U_{0} + U_{1} \cos (\Omega t + \varphi)$ periodic in time is induced 
uniformly over the cavity with the help of a top gate.
In the case of a slow potential, $\Omega \tau \ll T$, where $\tau$ is the time of one turn around the cavity, the (frozen \cite{MB04} ) scattering amplitude of a capacitor for an electron with incident energy $E$ and propagating in the linear edge state at time t is:

\begin{equation}
\label{08_01}
S^{SES}(t,E) = e^{ i\theta_{r} } \frac{ \sqrt{1 - T} - e^{i\phi(t,E)} }{ 1 - \sqrt{1 - T} e^{i\phi(t,E)} }\,.
\end{equation}
\ \\
\noindent
Here $\theta_{r}$ is the phase of the reflection amplitude $r = \sqrt{1-T}\, e^{i\theta_{r}}$ of the QPC connecting the circular edge state in the cavity to the linear edge state. $\phi(t,E) = \theta_{r} + \phi(E) - 2\pi eU(t)/\Delta$ is the phase accumulated by an electron with energy $E$ during one trip along the cavity and $\Delta$ is the level spacing in the cavity. 
The phase $\phi(E) = k_{F}L + (E - \mu) L/(\hbar v_{D})$ with $k_{F}$ a constant and $v_{D}$ a drift velocity can be taken to depend linearly on the energy. 
In the following, we consider the scattering amplitude for electrons with Fermi energy, $S^{SES}(t) \equiv S^{SES}(t,\mu)$. We are interested in the limit of a small transparency, $T \to 0$, when the width of the levels in the cavity is much smaller than the level spacing $\Delta$. The amplitude $U_{1}$ of the oscillating potential is chosen in such a way that during a period only one level of the cavity crosses the Fermi level $\mu$ in the linear edge state.
The time of crossing $t_{0}$ is defined by the condition $\phi(t_{0},\mu) = 0 \mod 2\pi$.
Introducing the deviation of the phase from its resonance value, $\delta\phi(t) = \phi(t,\mu) - \phi(t_{0},\mu)$, we obtain the scattering amplitude,

\begin{equation}
 \label{07_09}
S^{SES}(t) = - e^{i\theta_{r} }\, \frac{T + 2i\delta\phi(t) }{T - 2i\delta\phi(t)} + {\cal O} (T^{2}) \,.
\end{equation}
\ \\
\noindent
We keep only terms to leading order in $T \ll 1$.

There are two time moments when resonance conditions occur (two times of crossing).
The first crossing time is the instant when the level rises above the Fermi level and the second crossing time is when the level sinks below the Fermi level.
We denote these times $t_{0}^{(-)}$ and $t_{0}^{(+)}$, respectively.
At the time $t_{0}^{(-)}$ one electron is emitted by the cavity into the linear edge state, while at the time $t_{0}^{(+)}$ one electron enters the cavity, a hole is emitted.

We suppose that the constant part of the potential $U_{0}$ accounts for a detuning of the nearest electron level $E_{n}$ in the SES from the Fermi level.
Then the resonance times can be found from the following equation:

\begin{equation}
E_{n} + eU\left(t_{0}^{(\mp)}\right) = \mu_{0} \quad\Rightarrow\quad U_{0} + U_{1}\cos\left(\Omega t_{0}^{(\mp)} + \varphi \right) = 0\,.
\label{07_09_1}
\end{equation}
\ \\
\noindent
For $|eU_{0}| < \Delta/2$ and $|eU_{0}| < |eU_{1}| < \Delta - |eU_{0}|$ we find,

\begin{eqnarray}
t_{0}^{(\mp)} = \mp t_{0}^{(0)} \,-\, \dfrac{\varphi }{\Omega } \,,\quad t_{0}^{(0)} = \dfrac{1}{\Omega } \arccos\left(- \dfrac{U_{0}}{U_{1}} \right) .
\label{07_09_2}
\end{eqnarray}
\ \\
\noindent
The deviation from the resonance time, $\delta t^{(\mp)} = t - t_{0}^{(\mp)}$, can be related to the deviation from the resonance phase, $\delta\phi^{(\mp)} = \mp M \Omega \delta t^{(\mp)}$, where $\mp M= d\phi/dt|_{t = t_{0}^{(\mp)}}/\Omega = \mp 2\pi |e| \Delta^{-1} \sqrt{U_{1}^{2} - U_{0}^{2} }$.
With these definitions we can rewrite Eq.\,(\ref{07_09}) as follows:

\begin{equation}
S^{SES}(t) = e^{i\theta_{r}} \left\{
\begin{array}{ll}
\dfrac{t - t_{0}^{(+)} - i \Gamma_{0}}{t - t_{0}^{(+)} + i \Gamma_{0}}\,, & \left|t - t_{0}^{(+)} \right| \lesssim \Gamma_{0}\,, \\
\ \\
\dfrac{t - t_{0}^{(-)} + i \Gamma_{0}}{t - t_{0}^{(-)} - i \Gamma_{0}}\,, & \left|t - t_{0}^{(-)} \right| \lesssim \Gamma_{0}\,, \\
\ \\
1\,, & \left|t - t_{0}^{(\mp)} \right| \gg \Gamma_{0}\,.
\end{array}
\right.
\label{07_nsc_03_1} 
\end{equation}
\ \\
\noindent
Here $\Gamma_{0}$ is (half of) the time during which the level rises above or sink below the Fermi level:

\begin{eqnarray}
\Omega\, \Gamma_{0} = \dfrac{T\Delta}{ 4\pi |e| \sqrt{U_{1}^{2} - U_{0}^{2} }  } \,.
\label{nvd_22}
\end{eqnarray}
\ \\
\noindent
Eq. (\ref{07_nsc_03_1}) assumes that the overlap between the resonances is small,

\begin{equation}
\left| t_{0}^{(+)} - t_{0}^{(-)} \right|
 \gg \Gamma_{0}\,.
\label{07_09_3}
\end{equation}

\begin{figure}[t]
  \vspace{0mm}
  \centerline{
   \epsfxsize 8cm
   \epsffile{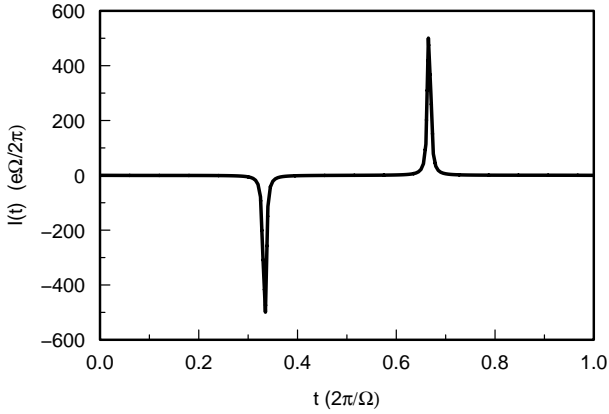}
             }
  \vspace{0mm}
  \nopagebreak
\caption{The time-dependent current, Eq.\,(\ref{07_05_20_01}), generated by the single-electron source at zero temperature. The positive (negative) peak corresponds to emission of an electron (a hole).  The parameters of the single electron source described by , Eq.\,(\ref{08_01}), are: $T = 0.1$, $U_{0} = 0.25 \Delta$, $U_{1} = 0.5 \Delta$, $\varphi = 0$.}
\label{figB1}
\end{figure}

The basic equation for the time-dependent current is 
(see, e.g., Ref.\,\onlinecite{BM06}),

\begin{equation}
I(t) = - \, \dfrac{ie }{2\pi } \int dE \left(-\, \dfrac{\partial f_{0} }{\partial E }\right) S^{SES}\, \dfrac{\partial \left( S^{SES}\right)^{*}}{\partial t } .
\label{07_05_20_01}
\end{equation}
\ \\
\noindent
Using Eq.\,(\ref{07_nsc_03_1}),  
we find the adiabatic current at zero temperature (for $0 < t < {\cal T}$):
\begin{equation}
\label{07_14}
I(t) \,=\, \frac{e}{\pi}\, \left\{ \dfrac{ \Gamma_{0}}{ \left( t - t_{0}^{(-)} \right)^2 + \Gamma_{0}^{2} } \,-\, \dfrac{\Gamma_{0}}{ \left( t - t_{0}^{(+)} \right)^2 + \Gamma_{0}^{2} } \right\} \,.
\end{equation}
\ \\
\noindent
In each time interval $2\pi/\Omega$ the current, shown in Fig.\,\ref{figB1}, consists of two pulses of Lorentzian shape with half-width $\Gamma_{0}$. The pulses correspond to the emission of an electron and a hole.
Integrating over time it is easy to check that the first pulse carries a charge $e$ while the second pulse carries a charge $-e$.

\section{Current correlation function}
\label{app2}

\subsection{General formalism}

Let the scatterer be connected via one-channel leads to reservoirs having different potentials,

\begin{equation}
V_{\alpha}(t) = V_{\alpha} + V_{\alpha}^{(\sim)}(t) \,.
\label{ac01}
\end{equation}
\ \\
\noindent
Following the approach developed in Refs.\,\onlinecite{JWM94, PB98}, we include the potential $V_{\alpha}^{(\sim)}(t) = V_{\alpha}^{(\sim)}(t + {\cal T})$, ${\cal T} = 2\pi/\Omega$,
oscillating with frequency $\Omega$ into the phase of the wave function for electrons injected into the circuit from reservoir ${\alpha}$.
The constant part of the potential changes the Fermi distribution function in contact ${\alpha}$, 

\begin{equation}
f_{\alpha}(E) = \dfrac{1}{1 + \exp{\dfrac{E - \mu_{\alpha} }{k_{B} T_{\alpha} } }  }\,, \quad \mu_{\alpha} = \mu_{0} + eV_{\alpha}\,.
\label{ac02}
\end{equation}
\ \\
\indent
We introduce the second quantization operator $\hat a^{\prime}_{\alpha}(E)$ annihilating an electron in the state with energy $E$ carrying a unit flux \cite{Buttiker92} in reservoir $\alpha$.
Then the corresponding distribution function is,

\begin{equation}
\left \langle a^{\prime\dag}_{\alpha}(E) a^{\prime}_{\alpha}(E^{\prime})  \right \rangle = f_{\alpha}(E) \delta \left( E - E^{\prime} \right)\,.
\label{ac03}
\end{equation}
\ \\
\noindent
If the reservoir $\alpha$ is subject to a periodic in time potential $V_{\alpha}^{(\sim)}(t)$, then the wave function for particles described by the operators $\hat a_{\alpha}^{\prime}$ is a Floquet type function having side-bands with energies $E_{n} = E + n\hbar\Omega$, $n = 0, \pm 1, \pm 2,\dots$
The amplitudes of side-bands are,

\begin{eqnarray}
\Upsilon_{\alpha,n} &=& \int\limits_{0}^{\cal T}  \dfrac{dt}{\cal T}\, e^{in\Omega t}\, \Upsilon_{\alpha}(t)\,,
\label{ac04} \\
\Upsilon_{\alpha}(t) &=& e^{-\,i\,  \frac{e}{\hbar} \int\limits_{-\infty}^{t} dt'\, V_{\alpha}^{(\sim)}(t')  } \,. \nonumber
\end{eqnarray}
\ \\
\indent
We suppose that there is no oscillating potential in the leads connecting the reservoirs to the scatterer.
Then the operator for particles in lead $\alpha$ is, \cite{PB98}

\begin{equation}
\hat a_{\alpha}(E) = \sum\limits_{n = -\infty}^{\infty} \Upsilon_{\alpha,n}\, \hat a_{\alpha}^{\prime}(E_{-n})\,.
\label{ac05}
\end{equation}
\ \\
\indent
If the scatterer is driven periodically then it is characterized by the Floquet scattering matrix $\hat S_{F}$.\cite{MB02}
We assume that the scatterer is driven with the same period ${\cal T}$ as the reservoirs.
The element $S_{F,\alpha\beta}(E_{n}, E)$ is a current scattering amplitude \cite{Buttiker92} for an electron incoming from the lead $\beta$ with energy $E$ to be scattered with energy $E_{n} = E + n\hbar\Omega$ into the lead $\alpha$.
With these amplitudes we find the operators for scattered particles, \cite{MB04}

\begin{eqnarray}
\hat b_{\alpha}(E) &=& \sum\limits_{\beta} \sum\limits_{m=-\infty}^{\infty} S_{F,\alpha\beta}(E, E_{m}) \hat a_{\beta}(E_{m}) \label{ac06} \\
&=& \sum\limits_{\beta} \sum\limits_{m=-\infty}^{ \infty} \sum\limits_{n=-\infty}^{\infty} S_{F,\alpha\beta}(E, E_{m}) \Upsilon_{\beta, n} \hat a_{\beta}^{\prime}(E_{m-n})\,.  \nonumber
\end{eqnarray}
\ \\
\indent
Now we calculate the symmetrized current correlation function in frequency representation,

\begin{equation}
P_{\alpha\beta}(\omega_{1},\omega_{2}) = \frac{1}{2} \left\langle \Delta \hat I_{\alpha}(\omega_1) \Delta \hat I_{\beta}(\omega_2) + \Delta \hat I_{\beta}(\omega_2)  \Delta \hat I_{\alpha}(\omega_1) \right\rangle\,,
\label{ac07}
\end{equation}
\ \\
\noindent
where $\left \langle \cdots  \right \rangle$ stands for quantum-statistical averaging over the (equilibrium) state of reservoirs,  $\Delta \hat I_{\alpha}(\omega) = \hat I_{\alpha}(\omega) - \left\langle \hat I_{\alpha}(\omega) \right\rangle$, and $\hat I_{\alpha}(\omega)$ is the operator for the current in lead $\alpha$,

\begin{equation}
\hat I_{\alpha}(\omega) = e \int\limits_{0}^{\infty} dE \left\{ \hat b_{\alpha}^{\dagger}(E) \hat b_{\alpha}(E+\hbar\omega) - \hat a_{\alpha}^{\dagger}(E) a_{\alpha}(E + \hbar\omega) \right\}\,.
\label{ac08}
\end{equation}
\ \\
\noindent
Using Eqs.\,(\ref{ac03}), (\ref{ac05}) -- (\ref{ac08}) we find

\begin{widetext}
\begin{eqnarray}
P_{\alpha\beta}(\omega_{1},\omega_{2})   & = & \sum\limits_{l = -\infty}^{\infty} 2\pi\, \delta(\omega_{1} + \omega_{2} - l\Omega )\, {\cal P}_{\alpha\beta, l}(\omega_{1})\,,
\nonumber \\
\label{ac09} \\
{\cal P}_{\alpha\beta, l}(\omega_{1}) &=& \dfrac{e^2}{h} \int dE\, \Bigg\{ \delta_{\alpha\beta}\, f_{\alpha\alpha}(E, E + \hbar\omega_{1}) \nonumber \\
&& - f_{\alpha\alpha}(E, E + \hbar\omega_{1}) \sum\limits_{n} \sum\limits_{ p,q} S_{F,\beta\alpha}\left(E_{l+n}\,, E_{p} \right) \Upsilon_{\alpha,p}\, S_{F,\beta\alpha}^{*} \left( E_{n} + \hbar\omega_{1}\,, E_{q} + \hbar\omega_{1} \right) \Upsilon_{ \alpha, q}^{*}  \nonumber  
\end{eqnarray}
\begin{eqnarray}
&& - f_{\beta\beta}(E, E + \hbar\omega_{2}) \sum\limits_{n} \sum\limits_{ p,q} S_{F,\alpha\beta}\left(E_{l+n}\,, E_{p} \right) \Upsilon_{\beta,p}\, S_{F,\alpha\beta}^{*} \left( E_{n} + \hbar\omega_{2}\,, E_{q} + \hbar\omega_{2} \right) \Upsilon_{ \alpha, q}^{*}  \nonumber \\
 \nonumber \\
&& + \sum\limits_{\gamma,\delta} \sum\limits_{n.m.s} \sum\limits_{p,q,p_{1},q_{1}} f_{\gamma\delta}\left(E_{n}\,, E_{m} + \hbar\omega_{1} \right)\, S_{F,\beta\gamma}\left(E_{l+s}, E_{n+q} \right) \Upsilon_{\gamma,q}\, S_{F,\alpha\gamma }^{*}\left(E, E_{n+p} \right) \Upsilon_{ \gamma,p }^{*}  \nonumber \\
\nonumber \\
&& \times S_{F,\alpha\delta}\left(E + \hbar\omega_{1}, E_{m+q_{1}} + \hbar\omega_{1} \right) \Upsilon_{\delta,q_{1}} S_{F, \beta\delta }^{*}\left(E_{s} + \hbar\omega_{1}, E_{m+p_{1}} +\hbar\omega_{1} \right) \Upsilon_{ \delta,p_{1} }^{*} \Bigg\} . \nonumber
\end{eqnarray}
\ \\
\noindent
Here
\begin{equation}
f_{\alpha\beta}(E_{1},E_{2}) = \dfrac{1}{2} \Big\{ f_{\alpha}(E_{1})\left[1 - f_{\beta}(E_{2}) \right] +  f_{\beta}(E_{2})\left[1 - f_{\alpha}(E_{1}) \right] \Big\}   \,. 
\end{equation}
\ \\
\indent
We are interested in the zero-frequency limit of the equation given above, when the noise can be conveniently represented as the sum of the thermal noise ${\cal P}_{\alpha\beta}^{(th)}$ (vanishing at $k_{B}T_{\alpha} =0$, $\forall \alpha$) and the shot noise ${\cal P}_{\alpha\beta}^{(sh)}$ (vanishing at $\Omega = 0$ and  $eV_{\alpha} = eV_{0}$, $\forall \alpha$).

\subsection{Zero frequency noise power}

At $l = 0$ and $\omega_{1} = \omega_{2} =0$ equation (\ref{ac09}) can be represented as follows:

\begin{subequations}
\label{ac10}
\begin{eqnarray}
{\cal P}_{\alpha\beta}  & = & \dfrac{e^2}{h} \int dE\, \left\{ {\cal P}_{\alpha\beta}^{(th)}(E) + {\cal P}_{\alpha\beta }^{(sh)}(E) \right\} \,,
\label{ac10a}
\end{eqnarray}
\begin{eqnarray}
{\cal P}_{\alpha\beta}^{(th)}(E) &=&  \delta_{ \alpha\beta } \left\{ f_{\alpha\alpha}(E, E) +  \sum\limits_{\gamma} F_{\alpha \gamma }( E) \right\} - F_{\alpha\beta}(E) - F_{\beta\alpha}(E)\,, \nonumber \\
\label{ac10b} \\
{\rm with} \quad\quad\quad
F_{\alpha\gamma}(E) &=& f_{\gamma\gamma }( E, E)  \sum\limits_{n,p,q}   S_{F,\alpha\gamma}\left(E_{n}, E_{q} \right) \Upsilon_{\gamma,q} S_{F,\alpha\gamma}^{ *}\left(E_{n}, E_{p} \right) \Upsilon_{ \gamma, p}^{*}\,,  \nonumber
\end{eqnarray}
\begin{eqnarray}
{\cal P}_{\alpha\beta}^{(sh)}(E) &=& \dfrac{1}{2} \sum\limits_{\gamma,\delta} \sum\limits_{n.m.s} \sum\limits_{p,q,p_{1},q_{1}} \Big\{ f_{\gamma}\left(E_{n} \right) - f_{\delta}\left(E_{m} \right) \Big\}^{2}   \nonumber \\
\label{ac10c} \\
&& \times S_{F,\beta\gamma}\left(E_{s}, E_{n+q} \right) \Upsilon_{\gamma,q}\, S_{F,\alpha\gamma }^{*}\left(E, E_{n+p} \right) \Upsilon_{ \gamma,p }^{*}\, S_{F,\alpha\delta}\left(E, E_{m+q_{1}} \right) \Upsilon_{\delta,q_{1}} S_{F, \beta\delta }^{*}\left(E_{s}, E_{m+p_{1}} \right) \Upsilon_{\delta,p_{1}}^{*} \,. \nonumber
\end{eqnarray}
\end{subequations}
\end{widetext}
\ \\
\indent
Now we show how Eq.\,(\ref{ac09}) was obtained.

\subsection{Derivation of the current correlation function}

To make the calculations more transparent it is convenient to represent the current as a sum, $\hat I_{\alpha}(\omega) = \hat I_{\alpha}^{(out)}(\omega) + \hat I_{\alpha}^{(in)}(\omega)$, of a current $I_{\alpha}^{(out)}$ carried by the scattered particles and a current $I_{\alpha}^{(in)}$ carried by the incident particles:

\begin{eqnarray}
\hat I_{\alpha}^{(out)}(\omega)   & = &  e \int\limits_{0}^{\infty} dE\,  \hat b_{\alpha}^{\dagger}(E) \hat b_{\alpha}(E+\hbar\omega)\,, \nonumber \\
\label{ap01} \\
\hat I_{\alpha}^{(in)}(\omega)   & = & -\, e \int\limits_{0}^{\infty} dE\,  \hat a_{\alpha}^{\dagger}(E) \hat a_{\alpha}(E+\hbar\omega)\,. \nonumber
\end{eqnarray}
\noindent
Then $P_{\alpha\beta}(\omega_{1},\omega_{2})$, Eq.\,(\ref{ac07}), can be represented as the sum of four terms,

\begin{equation}
\label{ap02}
P_{\alpha\beta}(\omega_{1},\omega_{2})    =  \sum\limits_{i,j = in, out}  P_{\alpha\beta}^{(i,j)}(\omega_{1},\omega_{2}) \,,
\end{equation}
\begin{equation}
P_{\alpha\beta}^{(i,j)}  =    \frac{1}{2} \left\langle \Delta \hat I_{\alpha}^{(i)}(\omega_1) \Delta \hat I_{\beta}^{(j)}(\omega_2) + \Delta \hat I_{\beta}^{(j)}(\omega_2)  \Delta \hat I_{\alpha}^{(i)}(\omega_1)\right\rangle . \nonumber
\end{equation}
\ \\
\noindent
We evaluate each of these four contributions separately.

\subsubsection{Correlator for incoming currents}

The first term in Eq.\,(\ref{ap02}) reads,

\begin{equation}
P_{\alpha\beta}^{(in,in)}(\omega_{1},\omega_{2}) = e^2 \iint\limits_{0}^{\infty} dE_{1}\, dE_{2} \, \frac{J_{\alpha\beta}^{(in,in)} + J_{\beta\alpha}^{(in,in)}}{2}  \,,
\label{ap03}
\end{equation}
\noindent
where

\begin{eqnarray}
J_{\alpha\beta}^{(in,in)} &=& \Big\langle \Big\{ \hat a_{\alpha}^{\dagger}(E_{1})\,\hat a_{\alpha}(E_{1} + \hbar\omega_{1}) \nonumber \\
&&- \Big\langle \hat a_{\alpha}^{\dagger}(E_{1})\,\hat a_{\alpha}(E_{1} + \hbar\omega_{1})\Big\rangle\Big\}  \nonumber \\
&\times&   \Big\{ \hat a_{\beta}^{\dagger}(E_{2})\,\hat a_{\beta}(E_{2} + \hbar\omega_{2}) \nonumber \\
&&- \left\langle \hat a_{\beta}^{\dagger}( E_{2})\,\hat a_{\beta}(E_{2} + \hbar\omega_{2}) \right\rangle  \Big\} \Big\rangle  .
\nonumber
\end{eqnarray}
\ \\
\noindent
In the correlation $J_{\beta\alpha}^{(in,in)}$ with the indices interchanged, the order of operators in each of the products contributing to $J_{\beta\alpha}^{(in,in)}$ is interchanged.
Using Wick's theorem, we represent the average of the product of four operators via the average of pair products and find,

\begin{eqnarray}
J_{\alpha\beta}^{(in,in)} &=& \Pi_{\alpha\beta}^{ (in,in)} \Xi_{\alpha\beta}^{(in,in)}\,, \nonumber \\
\label{ap05} \\
\Pi_{\alpha\beta}^{ (in,in)} &=& \Big\langle \hat a_{\alpha}^{\dagger}(E_{1})\,\hat a_{\beta}(E_{2}+\hbar\omega_{2}) \Big\rangle , \nonumber \\
\nonumber \\
\Xi_{\alpha\beta}^{(in,in)} &=& \left\langle \hat a_{\alpha}(E_{1}+\hbar\omega_{1})\,\hat a_{\beta}^{\dagger}( E_{2}) \right\rangle . \nonumber
\end{eqnarray}
\ \\
\noindent
Then using Eq.\,(\ref{ac05}) we obtain after straightforward but a little bit lengthy calculations,
\begin{eqnarray}
P_{\alpha\beta}^{(in,in)}\left(\omega_{1}, \omega_{2} \right)   & = & 2\pi\, \delta\left(\omega_{1} + \omega_{2} \right)\, {\cal P}_{\alpha\beta}^{(in,in)} \left(\omega_{1} \right) , \nonumber \\
\label{ap06} \\
{\cal P}_{\alpha\beta}^{(in,in)}\left(\omega_{1} \right) &=& \delta_{\alpha\beta}\, \dfrac{e^2}{h}\, \int dE_{1} f_{\alpha\alpha} \left( E_{1}\,, E_{1} + \hbar\omega_{1} \right) . \nonumber
\end{eqnarray}
\ \\
\noindent
This is exactly what could be expected for equilibrium electrons.
Therefore, uniform oscillating potentials at the reservoirs in themselves do not produce additional noise.

\subsubsection{Correlator between incoming and outgoing currents}

The next term in Eq.\,(\ref{ap02}) is,

\begin{eqnarray}
P_{\alpha\beta}^{(in,out)} = - e^2 \iint\limits_{0}^{\infty} dE_{1}\, dE_{2} \, \frac{J_{\alpha\beta}^{(in,out`)} + J_{\beta\alpha}^{(out,in)}}{2}  \,, \nonumber \\
\label{ap07}
\end{eqnarray}
\noindent
where

\begin{eqnarray}
J_{\alpha\beta}^{(in,out)} &=& \Pi_{\alpha\beta}^{ (in,out)} \Xi_{\alpha\beta}^{(in,out)}\,, \nonumber \\
\label{ap08} \\
\Pi_{\alpha\beta}^{ (in,out)} &=& \Big\langle \hat a_{\alpha}^{\dagger}(E_{1})\,\hat b_{\beta}(E_{2}+\hbar\omega_{2}) \Big\rangle , \nonumber \\
\nonumber \\
\Xi_{\alpha\beta}^{(in,out)} &=& \left\langle \hat a_{\alpha}(E_{1}+\hbar\omega_{1})\,\hat b_{\beta}^{\dagger}( E_{2}) \right\rangle . \nonumber
\end{eqnarray}
\ \\
\noindent
In the correlation $J_{\beta\alpha}^{(out,in)}$ the order of operators in the averages of pairs is interchanged.
Using Eqs.\,(\ref{ac05}) and (\ref{ac06}) we find,

\begin{eqnarray}
\Pi_{\alpha\beta}^{ (in,out)}   = \sum\limits_{n,m,p} \Upsilon_{\alpha,n}^{*} S_{F,\beta\alpha}\left( E_{2} + \hbar\omega_{2}\,, E_{2,m} + \hbar\omega_{2} \right) \nonumber \\
\times \Upsilon_{\alpha,p} f_{\alpha}(E_{1,-n}) \delta\left(E_{1,-n} - E_{2,m-p} -\hbar\omega_{2} \right) , \nonumber
\end{eqnarray}

\begin{eqnarray}
\Xi_{\alpha\beta}^{ (in,out)}    = \sum\limits_{n_{1},m_{1},p_{1}} \Upsilon_{ \alpha,n_{1} } \Upsilon_{\alpha,p_{1}}^{*}  \left[1 - f_{\alpha}\left(E_{1,-n_{1}} + \hbar\omega_{1}  \right) \right]  \nonumber \\
\times S_{F,\beta\alpha}^{*}\left( E_{2}\,, E_{2,m_{1}}\right)  \delta\left(E_{1,-n_{1}} - E_{2,m_{1}-p_{1}} +\hbar\omega_{1} \right) . \nonumber
\end{eqnarray}
\ \\
\noindent
Next we integrate over energy $E_{2}$ using the Dirac delta-function in $\Pi_{\alpha\beta}^{ (in,out)}$.
In the reminder we use $E_{2,m-p} = E_{1,-n} - \hbar\omega_{2}$ and find,

\begin{eqnarray}
P_{\alpha\beta}^{(in,out)}    = -\, \dfrac{e^{2} }{\hbar }  \int dE_{1} \sum\limits_{n,m,p} \sum\limits_{n_{1},m_{1},p_{1}} \nonumber \\
\nonumber \\
\times f_{\alpha\alpha} \left(E_{1,-n}\,, E_{1,-n_{1}} + \hbar\omega_{1} \right)
\Upsilon_{\alpha,n}^{*}  \Upsilon_{\alpha,p} \Upsilon_{\alpha,n_{1} } \Upsilon_{\alpha,p_{1} }^{*} \nonumber \\
\nonumber \\
\times \delta \left(\omega_{1} + \omega_{2} - \Omega \left[p - n - m -p_{1} + n_{1} + m_{1} \right] \right)  \nonumber \\
\nonumber \\
\times S_{F,\beta\alpha}\left( E_{1,p-n-m}\,, E_{1,p-n} \right) \nonumber \\
\nonumber \\
\times S_{F,\beta\alpha}^{*}\left( E_{1,p_{1} - n_{1} - m_{1} } + \hbar\omega_{1} \,, E_{1,p_{1} - n_{1} } + \hbar\omega_{1}\right) . \nonumber
\end{eqnarray}
\ \\
\noindent
We shift (under the integral over $E_{1}$): $E_{1} \to E_{1} + n\hbar\Omega$.
Then we introduce $w = n - n_{1}$ instead of $n_{1}$.
The sum over $w$ gives us $\delta_{w0}$.
Then we introduce $l = p - m - p_{1} + m_{1}$ instead of $m$ and $r = p_{1} - m_{1}$ instead of $m_{1}$.
Finally we get,

\begin{equation}
P_{\alpha\beta}^{(in,out)}\left(\omega_{1}, \omega_{2} \right)   =  \sum\limits_{l= -\infty}^{\infty} 2\pi\, \delta\left(\omega_{1} + \omega_{2} -l\Omega \right)\, {\cal P}_{\alpha\beta,l}^{(in,out)}  ,
\label{ap09}
\end{equation}
\begin{eqnarray}
{\cal P}_{\alpha\beta,l}^{ (in,out) } \left( \omega_{1} \right) = -\, \dfrac{e^2}{h}\, \int dE_{1} f_{\alpha\alpha} \left( E_{1}\,, E_{1} + \hbar\omega_{1} \right)  \nonumber \\
\nonumber \\
\times \sum\limits_{r,p,p_{1} } S_{F,\beta\alpha }\left( E_{1,l+r}\,, E_{1,p} \right) \Upsilon_{\alpha,p} \nonumber \\
\times S_{F,\beta\alpha}^{*}\left( E_{1,r} + \hbar\omega_{1} \,, E_{1,p_{1} } + \hbar\omega_{1}\right) \Upsilon_{\alpha,p_{1} }^{*} . \nonumber
\end{eqnarray}
\ \\
\indent
With similar steps we find that $P_{\alpha\beta}^{(out,in)}$ can be obtained from $P_{\alpha\beta}^{(in,out)}$ if one replaces: $\alpha \leftrightarrow \beta$, $E_{1} \leftrightarrow E_{2}$, and $\omega_{1} \leftrightarrow \omega_{2}$.
Therefore, from Eq.\,(\ref{ap09}) we immediately obtain,

\begin{equation}
P_{\alpha\beta}^{(out,in)}\left(\omega_{1}, \omega_{2} \right)   =  \sum\limits_{l= -\infty}^{\infty} 2\pi\, \delta\left(\omega_{1} + \omega_{2} -l\Omega \right)\, {\cal P}_{\alpha\beta,l}^{(out,in)}  ,
\label{ap12}
\end{equation}
\begin{eqnarray}
{\cal P}_{\alpha\beta,l}^{ (out,in) } \left( \omega_{2} \right) = -\, \dfrac{e^2}{h}\, \int dE_{2} f_{\beta\beta} \left( E_{2}\,, E_{2} + \hbar\omega_{2} \right)  \nonumber \\
\nonumber \\
\times \sum\limits_{r,p,p_{1} } S_{F,\alpha\beta }\left( E_{2,l+r}\,, E_{2,p} \right) \Upsilon_{\beta,p} \nonumber \\
\times S_{F,\alpha\beta}^{*}\left( E_{2,r} + \hbar\omega_{2} \,, E_{2,p_{1} } + \hbar\omega_{2}\right) \Upsilon_{\beta,p_{1} }^{*} . \nonumber
\end{eqnarray}
\ \\
\noindent
To compare subsequently Eqs.\,(\ref{ap09}) and (\ref{ap12}) with Eq.\,(\ref{ac09}) we need additionally to redefine: $r\to n$ and $p_{1} \to q$.

\subsubsection{Correlator between outgoing currents}

The last term in Eq.\,(\ref{ap02}) reads,

\begin{eqnarray}
P_{\alpha\beta}^{(out,out)} = - e^2 \iint\limits_{0}^{\infty} dE_{1}\, dE_{2} \, \frac{J_{\alpha\beta}^{(out,out)} + J_{\beta\alpha}^{(out,out)}}{2}  \,, \nonumber \\
\label{ap13}
\end{eqnarray}
\ \\
\noindent
where

\begin{eqnarray}
J_{\alpha\beta}^{(out,out)} &=& \Pi_{\alpha\beta}^{ (out,out)} \Xi_{\alpha\beta}^{(out,out)}\,, \nonumber \\
\nonumber \\
\Pi_{\alpha\beta}^{(out,out)} &=& \Big\langle \hat b_{\alpha}^{\dagger}(E_{1})\,\hat b_{\beta}(E_{2}+\hbar\omega_{2}) \Big\rangle , \nonumber
\nonumber \\
\Xi_{\alpha\beta}^{ (out,out)} &=& \left\langle \hat b_{\alpha}(E_{1}+\hbar\omega_{1})\,\hat b_{\beta}^{\dagger}( E_{2}) \right\rangle . \nonumber
\end{eqnarray}
\ \\
\noindent
In the correlation $J_{\beta\alpha}^{(out,out)}$ the order of operators in the pair averages is interchanged.

Using Eqs.\,(\ref{ac05}) and (\ref{ac06}) we calculate,

\begin{eqnarray}
\Pi_{\alpha\beta}^{ (out,out)}   = \sum\limits_{\gamma} \sum\limits_{n,m,p,q}  \delta\left( E_{1,n-p} - E_{2,m-q} -\hbar\omega_{2} \right) \nonumber \\
\times f_{\gamma}\left(E_{1,n-p} \right) S_{F,\alpha\gamma}^{*}\left( E_{1}\,, E_{1,n} \right)   \nonumber \\
\times  \Upsilon_{\gamma,p}^{*} S_{F, \beta \gamma }\left( E_{2} + \hbar\omega_{2}\,, E_{2,m} + \hbar\omega_{2} \right) \Upsilon_{\gamma,q}  \,, \nonumber
\end{eqnarray}

\begin{eqnarray}
\Xi_{\alpha\beta}^{ (out,out)}    = \sum\limits_{\gamma_{1}} \sum\limits_{n_{1}, m_{1}, p_{1}, q_{1}} \delta\left( E_{1,n_{1}-p_{1}} + \hbar\omega_{1} - E_{2,m_{1}-q_{1}} \right) \nonumber \\
\times \left[ 1 - f_{\gamma_{1}}\left( E_{1,n_{1}-p_{1}} + \hbar\omega_{1} \right) \right] S_{F,\beta\gamma_{1}}^{*} \left( E_{2}\,, E_{2,m_{1}} \right)  \nonumber \\
\times \Upsilon_{ \gamma_{1}, q_{1}}^{*} S_{F, \alpha\gamma_{1} }\left( E_{1} + \hbar\omega_{1}\,, E_{1, n_{1} } + \hbar\omega_{1} \right) \Upsilon_{\gamma_{1},p_{1}} \,. \nonumber
\end{eqnarray}
\ \\
\noindent
Then we integrate over energy $E_{2}$ using the Dirac delta-function in $\Pi_{\alpha\beta}^{ (out,out)}$.
In the rest we use $E_{2} = E_{1,n - p - m + q} - \hbar\omega_{2} = E_{1,n_{1} + q_{1} - p_{1} - m_{1}} + \hbar\omega_{1}$ and find,

\begin{eqnarray}
P_{\alpha\beta}^{(out,out)}    = \dfrac{e^{2} }{\hbar }  \int dE_{1} \sum\limits_{\gamma, \gamma_{1}} \sum\limits_{n,m,p,q} \sum\limits_{n_{1},m_{1},p_{1},q_{1}} \nonumber \\
\times f_{\gamma\gamma_{1}} \left(E_{1,n-p}\,, E_{1,n_{1}-p_{1}} + \hbar\omega_{1} \right)
\nonumber \\
\times \delta \left(\omega_{1} + \omega_{2} - \Omega \left[n+q-p-m - n_{1} - q_{1} + p_{1} + m_{1} \right] \right)  \nonumber \\
\times S_{F,\alpha\gamma}^{*}\left( E_{1}\,, E_{1,n} \right) \Upsilon_{\gamma,p}^{*} S_{F, \beta \gamma }\left( E_{1,n-p-m+q}\,, E_{1,n - p + q}\right)   \nonumber \\
\times S_{F,\beta\gamma_{1}}^{*} \left( E_{1, n_{1} + q_{1} - p_{1} - m_{1} } + \hbar\omega_{1}\,, E_{1,n_{1} + q_{1} - p_{1} } + \hbar\omega_{1}  \right) \nonumber \\
\times \Upsilon_{\gamma,q}  \Upsilon_{ \gamma_{1}, q_{1}}^{*} S_{F, \alpha\gamma_{1} }\left( E_{1} + \hbar\omega_{1}\,, E_{1, n_{1} } + \hbar\omega_{1} \right) \Upsilon_{\gamma_{1},p_{1}} \,. \nonumber
\end{eqnarray}
\ \\
\noindent
To simplify we introduce $t = n-p$ instead of $n$, $w = n_{1} - p_{1}$ instead of $n_{1}$, $l = n+q-p-m - n_{1} - q_{1} + p_{1} + m_{1}$ instead of $m$, and $s = n_{1} + q_{1} - p_{1} - m_{1} $ instead of $m_{1}$.
Then we get,

\begin{equation}
P_{\alpha\beta}^{(out,out)}\left(\omega_{1}, \omega_{2} \right)   =  \sum\limits_{l= -\infty}^{\infty} 2\pi\, \delta\left(\omega_{1} + \omega_{2} -l\Omega \right)\, {\cal P}_{\alpha\beta,l}^{(out,out)}  ,
\label{ap14}
\end{equation}
\begin{eqnarray}
{\cal P}_{\alpha\beta,l}^{ (out,out) } \left( \omega_{1} \right) = \dfrac{e^2}{h}\, \int dE_{1} \sum\limits_{\gamma, \gamma_{1}} \sum\limits_{s,t,w} \sum\limits_{p,q,p_{1},q_{1}} \nonumber \\
\times f_{\gamma\gamma_{1}} \left(E_{1,t}\,, E_{1,w} + \hbar\omega_{1} \right)
\nonumber \\
\times S_{F,\alpha\gamma}^{*}\left( E_{1}\,, E_{1,t+p} \right) \Upsilon_{\gamma,p}^{*} S_{F, \beta \gamma }\left( E_{1,l+s}\,, E_{1,t + q}\right) \Upsilon_{\gamma,q}  \nonumber \\
\times S_{F,\beta\gamma_{1}}^{*} \left( E_{1,s} + \hbar\omega_{1}\,, E_{1,w + q_{1} } + \hbar\omega_{1}  \right) \Upsilon_{ \gamma_{1}, q_{1}}^{*} \nonumber \\
\times S_{F, \alpha\gamma_{1} }\left( E_{1} + \hbar\omega_{1}\,, E_{1, w + p_{1} } + \hbar\omega_{1} \right) \Upsilon_{\gamma_{1},p_{1}} \,. \nonumber
\end{eqnarray}
\ \\
\noindent
To compare subsequently with Eq.\,(\ref{ac09}) we need additionally to redefine: $t\to n$, $w \to m$, $p_{1} \leftrightarrow q_{1}$, and $\gamma_{1} \to \delta$.

Collecting together equations (\ref{ap06}), (\ref{ap09}), (\ref{ap12}), and (\ref{ap14})  we arrive at Eq.\,(\ref{ac09}).

\subsection{Adiabatic regime}

In the adiabatic regime the Floquet scattering matrix elements to leading order in $\Omega\to 0$ are the Fourier coefficients for the frozen scattering matrix $\hat S(t,E)$, \cite{MB02}

\begin{equation}
S_{F,\alpha\beta}\left(E_{n}, E_{m}\right) = S_{ \alpha \beta, n-m} \left(E \right)\,.
\label{ac13}
\end{equation}
\ \\
\noindent
Within this approximation we find from Eqs.\,(\ref{ac10b}), (\ref{ac10c}) the following,

\begin{subequations}
\label{ac14}
\begin{equation}
\label{ac14a}
\begin{array}{r}
{\cal P}_{\alpha\beta}^{(th,ad)}(E) = - f_{\alpha \alpha }(E, E) \overline{\left| S_{\beta\alpha}(E) \right|^2}\\
 - f_{\beta \beta }(E, E) \overline{\left| S_{\alpha\beta}(E) \right|^2}  \\
\ \\
+ \delta_{ \alpha\beta } \left\{ f_{\alpha\alpha}(E, E) +  \sum\limits_{\gamma} f_{\gamma \gamma }(E, E) \overline{\left| S_{\alpha\gamma}(E) \right|^2} \right\}  \,,
\end{array}
\end{equation}
\begin{eqnarray}
{\cal P}_{\alpha\beta}^{(sh,ad)}(E) = \dfrac{1}{2} \sum\limits_{\gamma,\delta} \sum\limits_{ q = - \infty }^{\infty} \left\{f_{\gamma}(E_{q}) - f_{\delta}(E) \right\}^{ 2} \nonumber \\
\times \Phi_{\alpha, q}^{(\gamma\delta)}\, \Phi_{\beta, q}^{(\gamma\delta)\,*}\,, \quad \label{ac14b}
\end{eqnarray}
\end{subequations}
where $\Phi_{\alpha, q}$ is a Fourier transform of

\begin{equation}
\quad\quad \Phi_{\alpha}^{(\gamma\delta)}(t) =  S_{\alpha\gamma}^{*}(t,E) \Upsilon_{\gamma}^{*}(t) S_{\alpha\delta}(t,E) \Upsilon_{\delta}(t) .
\label{ac14c}
\end{equation}

\ \\
\noindent
Here the over bar stands for a time average, $\overline{X} = \int_{0}^{\cal T} dt X(t)/{\cal T}$.
Calculating the shot noise we made a shift of $E \to E - m\hbar\Omega$ and introduced $q = n-m $ instead of $m$.

One can see that the potentials oscillating at reservoirs have no effect on the thermal noise. Their effect on the shot noise in the adiabatic regime can be taken into account formally by changing the phase of the scattering elements $S_{\varphi\rho}(t,E)$ by the factor $\Upsilon_{\rho}(t)$, Eq.\,(\ref{ac04}).

\subsection{Zero-temperature adiabatic regime}

At zero temperatures there is no thermal noise.
Calculating the shot noise we take into account that in the adiabatic regime the frequency $\Omega$ is so small that we can neglect the energy dependence of the scattering matrix elements over the interval of order several $\hbar\Omega$. \cite{MB02}$^{,}$ \cite{MB04}
In addition we assume also that all the potential differences $V_{\alpha\beta} = V_{\alpha} - V_{\beta}$ are small compared to the significant energy scales of the scattering matrix. 
Then with Eq.\,(\ref{ac14b}) the integral over energy in Eq.\,(\ref{ac10a}) becomes trivial and we find,

\begin{equation}
{\cal P}_{\alpha\beta}^{(sh,ad)} = \dfrac{e^2 \Omega }{4\pi } \sum\limits_{\gamma,\delta} \sum\limits_{ q = - \infty }^{\infty} \left|\dfrac{eV_{\gamma\delta} }{\hbar\Omega} - q \right| \Phi_{\alpha, q}^{(\gamma\delta)} \, \Phi_{\beta, q}^{(\gamma\delta)\,*}\,.
\label{ac15_1}
\end{equation}
\ \\
\noindent
Note the dc bias and ac bias enter this equation in a strongly non-equivalent way.

\section{Probability description of the current cross-correlator for a circuit with SESs}
\label{app3}

The single-electron source emits electrons and holes which are uncorrelated. Hence electrons $(e)$ and holes $(h)$ contribute to noise independently, ${\cal P}_{12} = {\cal P}_{12}^{(e)} + {\cal P}_{12}^{(h)}$. 
In the adiabatic regime we can neglect the energy dependence of the scattering matrix.
Therefore, electrons and holes contribute to the noise equally, ${\cal P}_{12}^{(e)} = {\cal P}_{12}^{(h)} = 0.5 {\cal P}_{12}$.
Below we restrict ourself to the electron contribution.
We assume that the circuit has two inputs and two outputs $1$ and $2$. In each input  there is a SES emitting one electron per period.

\subsection{Classical versus quantum regimes}

It was noticed in Ref.\,\onlinecite{SMB09} that the cross-correlator ${\cal P}_{12}^{(e)}$ is related to the electron number correlator $\delta {\cal N}_{12}$ as follows,

\begin{equation}
{\cal P}_{12}^{(e)} \,=\, \dfrac{e^2 \Omega }{2\pi }\, \delta {\cal N}_{12} \,,
\label{07_02_10}
\end{equation}
\ \\
\noindent
where

\begin{equation}
\delta {\cal N}_{12} \,=\, {\cal N}_{12} \,-\, {\cal N}_{1} {\cal N}_{2} \,.
\label{07_02_08}
\end{equation}
\ \\
\noindent
Here ${\cal N}_{12}$ is the probability to find one electron in output $1$ and one electrons in output $2$ {\it during} the period ${\cal T}$, whereas ${\cal N}_{j}$ is the probability to find an electron in output $j = 1, 2$ during the same period.

To determine the probabilities entering Eq.\,(\ref{07_02_08}) we need to consider a specific circuit.
We consider the one given in Fig.\,\ref{fig2}.
For this circuit the quantum-mechanical amplitudes ${\cal A}_{ij}$ for an electron emitted by the source $j = L, R$ to arrive at the output $i = 1, 2$ are the following:

\begin{eqnarray}
{\cal A}_{1L} \,=\, e^{i k_{F} L_{1L}}\,  t_{L}\, r_{C}\,, \quad {\cal A}_{1R} \,=\, e^{i k_{F} L_{1R}}\,   t_{R}\,   t_{C}\,,\nonumber \\
\label{07_ppa_14} \\
{\cal A}_{2L} \,=\, e^{i k_{F} L_{2L}}\,  t_{L}\,   t_{C}\,, \quad {\cal A}_{2R} \,=\, e^{i k_{F} L_{2R}}\,  t_{R}\, r_{C}\,. \nonumber
\end{eqnarray}
\ \\
\noindent
With these amplitudes we find single-particle probabilities,

\begin{eqnarray}
{\cal N}_{1} &=& \left| {\cal A}_{1L} \right|^2 + \left| {\cal A}_{1R} \right|^2 = T_{L} + T_{C}\left(T_{R} - T_{L} \right) \,, \nonumber \\
\label{07_ppa_15} \\
{\cal N}_{2} &=& \left| {\cal A}_{2L} \right|^2 + \left| {\cal A}_{2R} \right|^2 = T_{L} - T_{C}\left(T_{R} - T_{L} \right) \,. \nonumber
\end{eqnarray}
\noindent
The calculation of the two-particle probability ${\cal N}_{12}$ depends crucially on whether electrons collide at the central QPC or not.

If electrons pass the QPC $C$ at different times, $\Delta t^{(-)} \gg \Gamma_{L}, \Gamma_{R}$, then there are two independent processes contributing to ${\cal N}_{12}$ with amplitudes ${\cal A}^{(2)}_{I} = {\cal A}_{1L} {\cal A}_{2R}$ and ${\cal A}^{(2)}_{II} = {\cal A}_{1R} {\cal A}_{2L}$.
Since the two-particle amplitudes factorize into the product of single-particle amplitudes, we term this the {\it classical} regime.
With these amplitudes we find,

\begin{eqnarray}
{\cal N}_{12} &=& \left| {\cal A}^{(2)}_{I} \right|^2 + \left| {\cal A}^{(2)}_{II} \right|^2  = T_{L} T_{R} \left(R_{C}^{2} \,+\, T_{C}^{2} \right) .
\label{07_ppa_17}
\end{eqnarray}
\ \\
\noindent
Using Eqs.\,(\ref{07_ppa_15}) and (\ref{07_ppa_17}) in Eq.\,(\ref{07_02_08}) we find the cross-correlator ${\cal P}_{12}^{(e)}$, Eq.\,(\ref{07_02_10}), to be the same as the one given in Eq.\,(\ref{n02_01}) (times $0.5$ to account for the electron contribution).

In contrast, if electrons can collide at the QPC $C$, $\Delta t^{(-)} = 0$, then the two particle amplitude is given by the Slater determinant,

\begin{equation}
{\cal A}^{(2)} \,=\, \det \left|
\begin{array}{ll}
{\cal A}_{1L} & {\cal A}_{1R} \\
\ \\
{\cal A}_{2L} & {\cal A}_{2R}
\end{array}
\right| .
\label{07_ppa_09}
\end{equation}
\ \\
\noindent
This is why we call this regime {\it quantum}.
Then the two-particle probability reads,

\begin{equation}
{\cal N}_{12} = \left| {\cal A}^{(2)} \right|^2 = T_{L} T_{R}\,.
\label{07_ppa_20}
\end{equation}
\ \\
\noindent
Note this equation is independent of the parameters of the central QPC, that can be used as an indication of a quantum regime. 
We emphasize that in the quantum regime the two-particle probability becomes the Glauber joint detection probability \cite{Glauber63}, since electrons after collision of the QPC $C$ arrive at the outputs $1$ an $2$ simultaneously (disregarding a possible difference in arrival times due to the different distances).
With Eqs.\,(\ref{07_ppa_20}), (\ref{07_ppa_15}), (\ref{07_02_08}), and (\ref{07_02_10}) we recover the result given in Eq.\,(\ref{n02_02}).

\subsection{Positive two-particle correlations in quantum regime}

Let us show that in the quantum regime colliding electrons are positively correlated.
To this end we represent the single-particle probabilities as the sum of contributions due to each of sources, ${\cal N}_{i} = {\cal N}_{i}^{(L)} + {\cal N}_{i }^{(R)}$ with ${\cal N}_{i}^{(j)} = \left| {\cal A}_{i j} \right|^2$ , $i = 1, 2$, $j = L, R$.
Then we split the particle number correlator $\delta {\cal N}_{12}$, Eq.\,(\ref{07_02_08}), into the sum of three contributions,

\begin{equation}
\delta {\cal N}_{12} = \delta {\cal N}_{12}^{(LL)} + \delta {\cal N}_{12}^{(RR)} + \delta {\cal N}_{12}^{(\widehat{LR})} \,.
\label{n06}
\end{equation}
\ \\
\noindent
Here the first two terms are contributions due to either source alone, $\delta {\cal N}_{12}^{(jj)} = - {\cal N}_{1}^{(j)} {\cal N}_{2}^{(j)}$, $j = L, R$.
Since the source emits single particles this contribution to the cross-correlator $\delta {\cal N}_{12}$ is definitely negative.
The third contribution is due to a joint action of both sources,

\begin{equation}
\delta {\cal N}_{12}^{(\widehat{LR})} = {\cal N}_{12} - {\cal N}_{1}^{(L)}{\cal N}_{2}^{(R)} - {\cal N}_{1}^{(R)}{\cal N}_{2}^{(L)} \,.
\label{n07}
\end{equation}
\ \\

In the classical regime we use Eq.\,(\ref{07_ppa_17}) and find, $\delta {\cal N}_{12}^{(\widehat{LR})} = 0$, i.e., the particles emitted by different sources remain uncorrelated.
In contrast in the quantum regime, using Eq.\,(\ref{07_ppa_20}), we get,

\begin{equation}
\delta {\cal N}_{12}^{(\widehat{LR})} = 2 T_{L} T_{R} R_{C} T_{C}\,.
\label{07_ppa_21}
\end{equation}
\ \\
\noindent
Therefore, in this regime the particles emitted by two sources and colliding at the central QPC $C$, see Fig.\,\ref{fig2}, become positively correlated. 
We stress the total overall correlation  ${\cal N}_{12}$ remains negative.


\begin{thebibliography}{11}

\bibitem{Feve07}
G. F\`{e}ve
A. Mah\'e, J.-M. Berroir, T. Kontos, B. Pla\c{c}ais, D. C. Glattli, A. Cavanna, B. Etienne, Y. Jin,
Science {\bf 316}, 1169 (2007).

\bibitem{OSMB08}
S. Ol'khovskaya, J. Splettstoesser, M. Moskalets, and M. B\"uttiker,
Phys. Rev. Lett. {\bf 101}, 166802 (2008).

\bibitem{SMB09}
J. Splettstoesser, M. Moskalets, and M. B\"uttiker,  Phys. Rev. Lett. {\bf 103}, 076804 (2009).

\bibitem{Blumenthal07}
M. D. Blumenthal,
B. Kaestner, L. Li, S. Giblin, T. J. B. M. Janssen, M. Pepper, D. Anderson, G. Jones, D. A. Ritchie, Nature Physics {\bf 3}, 343 (2007). 

\bibitem{Wright08}
S. J. Wright, M. D. Blumenthal, Godfrey Gumbs,  A. L. Thorn, M. Pepper,   T. J. B. M. Janssen, S. N. Holmes, D. Anderson, G. A. C. Jones, C. A. Nicoll, and D. A. Ritchie, Phys. Rev. B. {\bf 78}, 233311 (2008).

\bibitem{YS92}
M. B\"{u}ttiker, Phys. Rev. Lett. 68, 843, (1992);
B. Yurke and D. Stoler, Phys. Rev. A {\bf 46}, 2229 (1992).

\bibitem{SSB04}
P. Samuelsson, E. V. Sukhorukov, and M. B\"uttiker, Phys. Rev. Lett. {\bf 92}, 026805 (2004).

\bibitem{SNB09}
P. Samuelsson, I. Neder, and M. B\"{u}ttiker, Phys. Rev. Lett. {\bf 102}, 106804  (2009);
Proceedings of the Nobel Symposium 2009, Qubits for future quantum computers, May 2009 in Goteborg, Sweden, Phys. Scr. T {\bf 137},  014023 (2009).

\bibitem{Neder07}
I. Neder,
N. Ofek, Y. Chung, M. Heiblum, D. Mahalu, V. Umansky, Nature {\bf 448}, 333 (2007).

\bibitem{Mandel99}
L. Mandel, Rev. Mod. Phys. {\bf 71}, S274 (1999).

\bibitem{Loudon91}
R. Loudon, in: J.A. Blackman, J. Taguena (Eds.), Disorder in Condensed Matter Physics, Clarendon Press, Oxford, 1991, p. 441.

\bibitem{BB00}
Ya. M. Blanter and M. B{\" u}ttiker,
Physics Reports {\bf 336},  1 (2000).

\bibitem{Buttiker90}
M.  B\"{u}ttiker, Phys. Rev. Lett. {\bf 65}, 2901 (1990).

\bibitem{Samuelsson06}
P.  Samuelsson and M. B\"{u}ttiker, Phys. Rev. B {\bf 73}, 041305 R (2006) .

\bibitem{Beenakker06}
C. W. J. Beenakker, in "Quantum Computers, Algorithms and  Chaos", International School of Physics Enrico Fermi, Vol. 126 IOS Press, Amsterdam, 2006; See also cond-mat/0508488.

\bibitem{HOM87}
C. K. Hong, Z. Y. Ou, and L. Mandel, Phys. Rev. Lett. {\bf 59}, 2044 (1987).

\bibitem{note1}
There is an interesting proposal how to produce separate electron and hole streams into edge states : F. Batista, P. Samuelsson, arXiv:1006.0136 (unpublished).

\bibitem{ILL97}
D. A. Ivanov, H.-W. Lee, and L. S. Levitov, Phys. Rev. B {\bf 56}, 6839 (1997)

\bibitem{KKL06}
J. Keeling, I. Klich, and L. S. Levitov, Phys. Rev. Lett. {\bf 97}, 116403 (2006).

\bibitem{ZSAM09}
J. Zhang, Y. Sherkunov, N. d�Ambrumenil, and B.~Muzykantskii,
Phys. Rev. B {\bf 80}, 245308 (2009).

\bibitem{FDJ08}
G. F\'{e}ve, P. Degiovanni, and Th. Jolicoeur, Phys. Rev. B {\bf 77} 035308 (2008).

\bibitem{DGF09}
P. Degiovanni, Ch. Grenier, and G. F\'eve, Phys. Rev. B {\bf 80}, 241307(R) (2009).

\bibitem{BM10}
M. B\"{u}ttiker and M. Moskalets,
Int. J. of Mod. Phys. B {\bf 24}, 1555 (2010).

\bibitem{vrmpmb} 
V. S. Rychkov, M. L. Polianski, and M. B\"uttiker, Phys. Rev. B {\bf 72}, 155326 (2005).

\bibitem{mmmb} 
M. Moskalets and M. B\"uttiker, Phys. Rev. B {\bf 70}, 245305 (2004).

\bibitem{reydellet} 
L.-H. Reydellet, P. Roche, D. C. Glattli, B. Etienne, and Y. Jin, Phys. Rev. Lett. {\bf 90}, 176803 (2003).
 
\bibitem{SOMB08}
J. Splettstoesser,
S. Ol'khovskaya, M. Moskalets, and M. B\"uttiker,
Phys. Rev. B {\bf 78}, 205110 (2008).

\bibitem{PTB96}
A. Pr\^{e}tre, H. Thomas, and M. B\"{u}ttiker, Phys. Rev B. {\bf 54}, 8130 (1996).

\bibitem{Gabelli06}
J. Gabelli, G. F\`{e}ve, J.-M. Berroir, B. Pla\c{c}ais, A. Cavanna, B. Etienne, Y. Jin, D.C. Glattli, Science {\bf 313}, 499 (2006).

\bibitem{MSB08}
M. Moskalets, P. Samuelsson, and M. B\"{u}ttiker, Phys. Rev. Lett. {\bf 100}, 086601 (2008).

\bibitem{MB04}
M. Moskalets and M. B\"{u}ttiker, Phys. Rev. B {\bf 69}, 205316 (2004).

\bibitem{BM06}
M. B\"uttiker and M. Moskalets, Lecture Notes in Physics, {\bf 690}, 33 (2006).

\bibitem{JWM94}
A.-P. Jauho, N. S. Wingreen, and Y. Meir, Phys. Rev. B. {\bf 50}, 5528 (1994).

\bibitem{PB98}
M. H. Pedersen and M. B\"uttiker, Phys. Rev. B. {\bf 58}, 12993 (1998).

\bibitem{Buttiker92}
M. B\"{u}ttiker, Phys. Rev. B. {\bf 46}, 12485 (1992).

\bibitem{MB02}
M. Moskalets and M. B\"uttiker, Phys. Rev. B  {\bf 66}, 205320 (2002).


\bibitem{Glauber63} G. J. Glauber, Phys. Rev. {\bf 130}, 2529 (1963).



\end{thebibliography}
\end{document}